\documentclass{article}
\usepackage{times}
\usepackage{epsfig,amssymb,amsfonts,amsmath}
\usepackage{mathrsfs}
\usepackage{graphicx}
\usepackage{dcolumn}
\usepackage{bm}

\begin{document}

\title{Self organized criticality in an improved Olami-Feder-Christensen model}
\author{Gui-Qing Zhang$^{1,} $\thanks{nkzhanggq@163.com,zhanggq@mail.nankai.edu.cn}$\,$,
 Ugur Tirnakli$^{2,} $\thanks{ugur.tirnakli@ege.edu.tr}$\,$, 
 Lin Wang$^3$ and Tian-Lun Chen$^4$ \\ \\
$^1$ Department of Physics, Nanjing Forestry University, \\ 
Nanjing 210037, P.R.China \\
$^2$ Department of Physics, Faculty of Science, Ege University, \\ 
35100 Izmir-Turkey \\
$^3$ Adaptive Networks and Control Lab, \\ 
Department of Electronic Engineering, \\
Fudan University, Shanghai 200433, P.R.China \\
$^4$ Department of Physics, Nankai University, \\ 
Tianjin 300071, P.R.China}

\date{\today}

\maketitle

\begin{abstract}
{An improved version of the Olami-Feder-Christensen model has been introduced to
consider avalanche size differences. Our model well demonstrates the
power-law behavior and finite size scaling of avalanche size distribution
in any range of the adding parameter $p_{add}$ of the model.
The probability density functions (PDFs) for the avalanche size
differences at consecutive time steps (defined as returns) appear to be
well approached, in the thermodynamic limit, by $q$-Gaussian shape with appropriate
$q$ values which can be obtained a priori from the avalanche size exponent $\tau$.
For the small system sizes, however, return distributions are found to be consistent
with the crossover formulas proposed recently in Tsallis and Tirnakli,
J. Phys.: Conf. Ser. $\mathbf{201}$, 012001 (2010).
Our results strengthen recent findings of Caruso {\it et al.} [Phys. Rev. E $\mathbf{75}$,
055101(R) (2007)] on the real earthquake data which support the hypothesis that
knowing the magnitude of previous earthquakes does not make the magnitude
of the next earthquake predictable. Moreover, the scaling relation of the waiting time
distribution of the model has also been found.}
\end{abstract}


\newpage

\baselineskip=16pt
\section{Introduction}
Self-organized criticality (SOC) is a concept designed to describe
extended dynamical systems reaching a statistically stationary
state, characterized by power-law distribution functions in both
space and time, without any "fine tuning" of an external parameter.
SOC was first introduced as a subject by Bak, Tang and Wiesenfeld (BTW)
in 1987 \cite{PRL59}. In their well-known paper, they
proposed a sandpile model and found the system showed SOC phenomenon
with bulk conservation law and open boundary
conditions \cite{PRL59,PRL38}. SOC has been proposed as a way to
model the widespread occurrence of power laws, 
i.e., the abundance of long-range correlations in space and time 
in various systems, such as chemical reactions, evolution,
avalanches, forest burns, heart attacks, market crushes,
earthquakes, etc \cite{Cambridge,BAK}. In order to forecast
earthquakes, several statistical models of earthquakes embodying
such SOC features 
have been proposed and studied \cite{SOC,PRL68,Eur,PRE75,PRA461992}.
For example, one is the Burridge-Knopoff (BK) model \cite{SOC}, in
which an earthquake fault is modeled as an assembly of blocks
mutually connected via elastic springs which are slowly driven by
external force. Another extensively studied statistical model  might
be the Olami-Feder-Christensen (OFC) model, which was first
introduced by Olami, Feder and Christensen in 1992 as a
simplification of the BK model. Mapping the BK model into a
two-dimensional lattice, they simulated the earthquake behavior and
introduced dissipation into the family of the SOC
systems \cite{PRE48,PRL68,PH.D}. Numerical studies have revealed that
the OFC model exhibits apparently critical properties such as the
Gutenberg-Richter(GR) law and the Omori law \cite{GEO9}. For these
reasons, the OFC model has been regarded as a typical
nonconservative model exhibiting SOC.

Many works of OFC model have focused on the homogeneous lattice
network \cite{PRL77,PRL88,PRA46}, however, the actual transmission of
seismic energy or force is often
inhomogeneous \cite{EPJB,PHYSICA337}. We know that earthquakes occur
as a result of the relative motion of tectonic plates and the
seismic energy will be released in the form of earthquake
waves (primary wave or secondary wave). This process takes place from
the epicenter, which is below the earth surface and spread
through the elastic vibration of the rocks. Due to different
geological conditions, the earthquake wave in the rock will spread
with different velocities and rates of decay. This will cause
different energy decay in different geological conditions, therefore
the heterogeneity of energy transfer occurs. So it is reasonable to
assume that the real earthquake system is heterogeneous, and people
can easily conclude that the heterogeneous factor should be
investigated in the earthquake model. Recently, some works have
already been carried out along these lines: Baiesi and Paczuski
proposed a metric to quantify correlations between earthquakes based
on scale-free networks. According to this metric, typical events are
strongly correlated to only one or a few preceding ones \cite{PRE69}.
Thus a classification of events as foreshocks, main shocks, or
aftershocks emerges automatically. Epicenter network of OFC model
has been investigated by Peixoto and Prado \cite{PRE77}, in which
they obtain a direct network and show a sharp difference between
the conservative and nonconservative regimes. In the scale free and
directional network models, the energy is released either randomly
or uniformly. In contrast to them, our energy release relates to the
nature of adjacent rocks. We notice that the tectonic plates which
have higher stress are prone to be affected by other plates. It can
collect more energy or force released by other plates. In order to
simulate this phenomenon, we introduce edge weight which determines
how the energy is transferred from one point to another in the
coupled-map lattice, to investigate the SOC behavior on the
inhomogeneous network. This work aims to study the self-organized
criticality behavior of the non-conservative improved OFC model.

\section{The Model}
\textit{Original OFC model}. In the OFC model, ``stress`` variable
$F_{i}$ $(F_{i}\geq 0)$ is assigned to each site on a square lattice
with $L\times L$ sites. Initially, a random value in the interval
$[0,1]$ is assigned to each $F_{i}$, where $F_{i}$ is increased
with a constant rate uniformly over the lattice until, at a
certain site $i$, the $F_{i}$ value reaches the threshold,
$F_{th}=1$. Then, the site $i$ "topples" and a fraction of stress
$\alpha F_{i}$ $(0<\alpha<0.25)$ is transmitted to each of its four
nearest neighbors, while $F_{i}$ itself is reset to zero, namely,

\begin{eqnarray}
F_{i}\geq F_{th}\Rightarrow\left\{\begin{array}{lll}
&F_{i}\rightarrow 0,&\\
&F_{nn}\rightarrow F_{nn} +\alpha F_{i},&
\end{array}\right.
\end{eqnarray}
where "$nn$" denotes the set of nearest-neighbor sites of $i$. If
the stress of one "$nn$" site $j$ exceeds the threshold, i.e.,
$F_{j\in nn}\geq F_{th}=1$, the site $j$ also topples,
distributing a fraction of stress $\alpha F_{j}$ to its four
nearest neighbors. Such a sequence of topplings continues until
the stress of all sites on the lattice becomes smaller than the
threshold $F_{th}$. A sequence of toppling events, which is
assumed to occur instantaneously, corresponds to one seismic event
or an avalanche. After an avalanche, the system goes into an
interseismic period where uniform loading of $F$ is resumed, until
some of the sites reach the threshold and the next avalanche
starts. The transmission parameter $\alpha$ measures the extent of
nonconservation of the model. The system is conservative for
$\alpha=0.25$, and is nonconservative for $\alpha<0.25$.

\textit{Improvement on original OFC model}. It has been widely
accepted that earthquakes occur as a result of the relative motion
of tectonic plates. The plates move relatively to one another,
resulting in the build up of stress at the plate boundaries. When
the stress at the plate boundaries reaches to a level that cannot be
supported by friction between the plates, the strain energy is
released intermittently, that is, an earthquake happens. We notice
that the tectonic plates which bear higher stress are prone to be
affected by other plates. It can collect more energy or force
released by other plates. So it is reasonable that the plate with
higher stress will get more energy or force when its adjacent plate
is released. Based on the argument above, we can assume the edge
weight $w_{ij}(t)=\displaystyle\left[F_{i}(t)+F_{j}(t)\right]/2$, for the
simplicity of our model, which is determined by the seismogenic
forces of the two connected sites. This assumption is not only a
good simulation of the above, but also more importantly, it can be
used to model the heterogeneity of energy transfer. In order to
study the dynamics of our weighted OFC model, we should reconsider
the redistribution rule. Compared with the original OFC model, we
just need a new transmission parameter $\alpha_{j}$ defined as
below \cite{PRL88,PHYSICAZHANG}:

\begin{eqnarray}
\alpha\rightarrow\alpha_{j}(t)=a\times\displaystyle\frac{w_{ij}(t)}
{\sum\limits_{j\in nn}w_{ij}(t)}=a\times
\displaystyle\frac{F_{i}(t)+F_{j}(t)}{4\,F_{i}(t)+
\sum\limits_{j\in nn}F_{j}(t)}.
\end{eqnarray}
In our improved model, the factor $\alpha_{j}(t)$ defines the
level of local conservation of the system and can be adjusted by
parameter $a$. Therefore, for the sake of convenience, we
consider the parameter $a$ as the control parameter. For a generic initial
condition, the weighted OFC model, after some transients (discarded
in each run), builds up long-range spatial correlations, reaches a
critical state and generate a time series of avalanche size
{$S_{i}$}, $i = 1, . . . , n$. In particular, we will analyze a
time series of $n=10^{7}$ events.

\section{Simulation Results}

\subsection{Avalanche-Size Distributions--Effect of the control parameter}
In this section, we mainly analyze the probability distribution of
the avalanche sizes. The weighted OFC model generates an avalanche size
sequence and the avalanche size distribution is the frequency of the
occurrence of the avalanches with the same size. In our model,
there are a number of adjustable parameters. For example, we can adjust
the threshold for each node, so nodes can be considered as special
(this issue will be given in a future work) or one can also consider
the impact of network structure (this issue will be addressed below).
In this section, we consider the relationship between avalanche size
distribution and the control parameter $a$.
As in original OFC model, the control parameter can also be
used to measure avalanche behavior.
The continuous, nonconservative weighted OFC model exhibits SOC
behavior for a wide range of $a$ values. The avalanche size exponent
$\tau$ depends on $a$ \cite{PHYSICAZHANG}.

\begin{figure}[ht]
 \scalebox{0.65}{\includegraphics{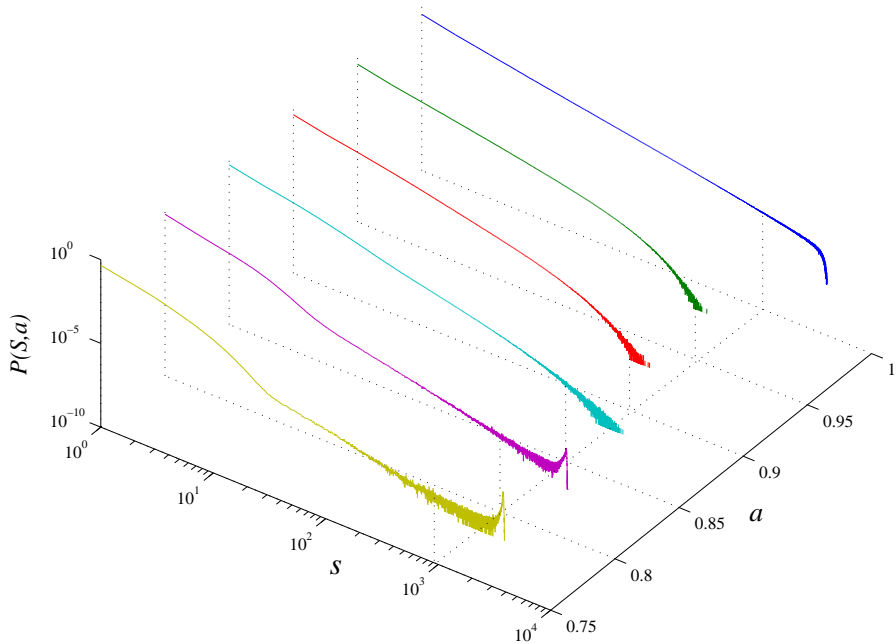}}
\caption{Distribution of avalanche sizes for different control
parameter $a$, $L=64$ and $p_{add}=0$. 
}
\end{figure}

In Fig.~1 we plot the avalanche size distribution for the weighted
OFC model with different control parameters $a$ \cite{NATUREPHYSICS}.
From this figure, we find that the system develops an approximate power-law
distribution for avalanche sizes 
in a wide range of parameter $a$.
When $a\leq 0.88$, the model only produces a power-law distribution
for avalanche sizes. The system not only shows power-law behavior
but also satisfies the finite-size scaling in the parameter
range $a=0.88$ to $a=1$.

\subsection{Avalanche-Size Distribution--finite-size scaling}
To verify the criticality of our weighted model, we study the
effect of increasing the system size $L$. We observe that,
for each constant value of $a$, the avalanche size exponent $\tau$
does not change, while the cutoff in the energy distribution scales
with the system size. Our weighted OFC model not only shows
power-law behavior in the avalanche size distribution, but also
satisfies the finite-size scaling behavior in the parameter range
mentioned above. In this part, we propose a simple finite-size scaling
analysis for the avalanche size distribution of the form

\begin{eqnarray}
P(S,L)&\propto& L^{-\beta}g(S/L^{\nu}) \,\, ,
\end{eqnarray}
where $g$ is the so-called universal scaling function, parameters
$\beta$ and $\nu$ are critical exponents used to
characterize scaling properties. $\nu$ may reflect the scaling
relationship between the cut off of the distribution function and
the system size, while $\beta$ is a normalization parameter.
Fig.~2 displays $P(S,L)$ versus the avalanche size $S$ for the weighted
OFC model on square lattice of size $L=32,48,64$ with control parameter
$a=1$ and the inset of Fig.~2 displays the transformed avalanche size
distribution, $L^{\beta}P(S,L)$, versus rescaled avalanche size,
$S/L^{\nu}$. A clear data collapse is evident for the proposed scaling
function with $\beta=2.456$, $\nu=2.002$.
The value of critical avalanche size exponent
($\tau=1.220\pm0.003$) \cite{PHYSICAZHANG} is in agreement
with the finite-size scaling hypothesis since for asymptotically
large $N$, it is well-known that $P(S)\sim S^{-\tau}$ with $\tau =\beta /
\upsilon$ \cite{IMPERIAL}, which gives $\tau=1.227$ for the obtained
values of $\beta$ and $\nu$. So far, we can conclude that our
weighted OFC model are not only self-organized but also critical.

\begin{figure}[ht]
\centering \scalebox{1.2}{\includegraphics{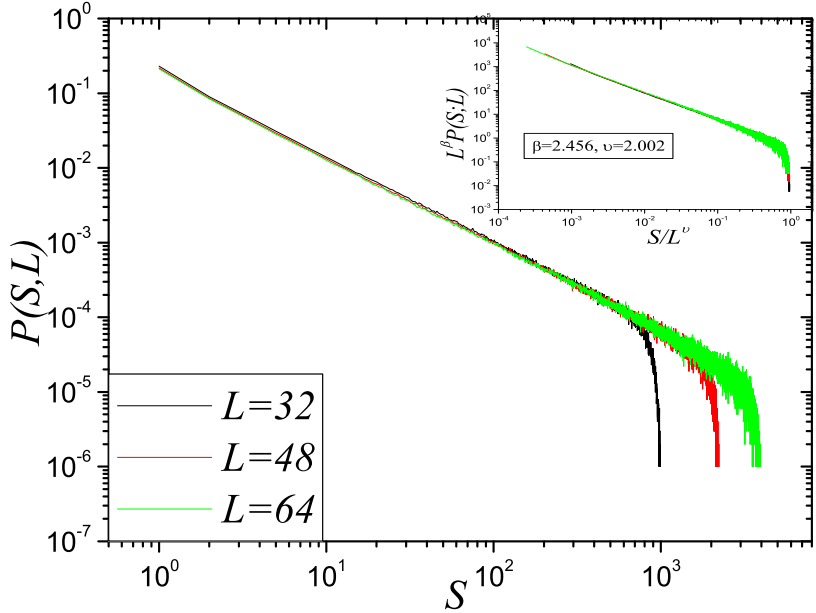}}
\caption{The avalanche size distribution $P(S,L)$ for the weighted
OFC model on square lattice with system size $L=32,48,64$.
In the inset, the transformed avalanche size distribution versus
rescaled avalanche size is given.}
\end{figure}

\subsection{Avalanche-Size Distribution--Effect of long range parameter}
In this section, we mainly discuss the effects of long-range
connectivity. The reasons why we introduce long range to our model
must be given first. To begin with, constructing networks from real
seismic data, Baiesi and Paczuski as well as Abe and Suzuki reported
the discoveries of the scale-free and the small-world features in
real earthquakes \cite{Eur,PHYSICA337,PRE69}. Then, according to the
geophysics and geology, heterogeneous character of real earthquake systems
and the effect of long-range interactions in real earthquakes were
found by Mori and Kawamura \cite{PRE772008}, for instance in
earthquake triggering and interaction, where the static stress may
involve relaxation processes in the asthenosphere with relevant
spatial and temporal long-range effects. Here, we introduce a small
fraction of long-range links (denoted as the long range parameter $p_{add}$)
in the lattice so as to obtain a small world topology.
The long-range connections largely reduce the average distance of the
original network (here, our model is based on the $NW$ small-world model).

In \cite{PHYSICAZHANG}, the effects of the control parameter have
mainly discussed, here we only discuss the behavior of the critical state.
In this model, the state of the system is controlled by the control
parameter and long rang parameter $p_{add}$. So, whether the system is in
self-organized criticality or not depends on these two parameters.
Depending on the long-range parameter, the network can produce a
rich repertoire of behaviors. In Fig.~3, we fix $a=1$ and show
examples of avalanche size distributions for various values of
$p_{add}$. For small values of $p_{add}$ ($p_{add}\leq0.3$), critical
avalanche size distributions are observed. This regime is
characterized by an approximate power-law distribution for avalanche
sizes almost up to the system sizes where an exponential cutoff is
observed. For larger values of $p_{add}$ ($p_{add}>0.3$), the
distribution is supercritical, that is, a substantial fraction of
triggering events spread through the whole system \cite{NATUREPHYSICS}.
When the control parameter equals to other values, the system shows
self-organized criticality behavior which is different from the behavior
above, and this can be explained as it is on the more susceptive critical state
($a=1$) than others.
In the inset of Fig.~3, we have simulated this behavior based on
different lattices and found that they show the same behavior
independent of lattice size expect for $L\leq 16$, which can be considered
as the effects of boundary conditions. 


\begin{figure}[ht]
 \scalebox{1.2}{\includegraphics{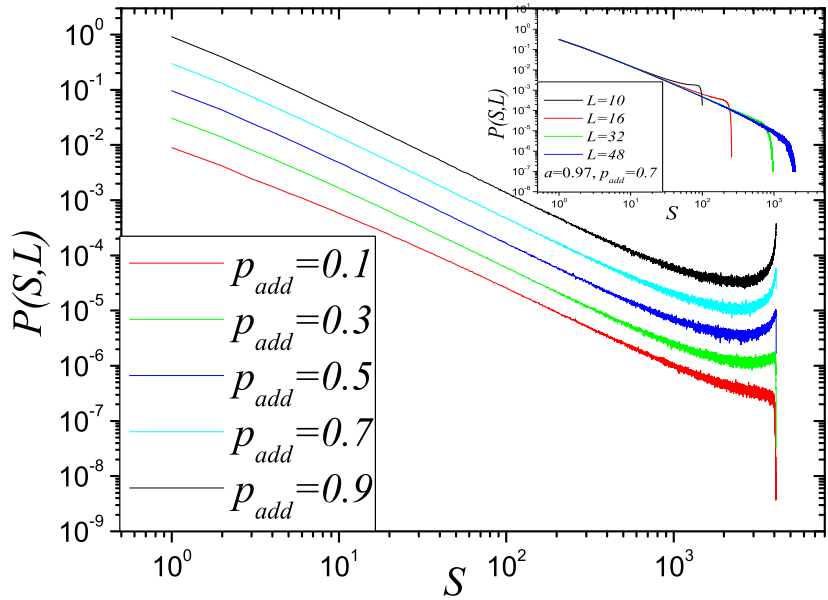}}
\caption{Avalanche size distributions for different long range
parameter $p_{add}$. The long range parameter near $p_{add}\leq0.3$
seems critical. For $p_{add}>0.3$, the distribution is supercritical.
The inset shows $P(S,L)$ versus the avalanche size $S$ for the weighted
OFC model on square lattice with system sizes $L=10,16,32$ and 48
for $a=0.97$ and $p_{add}=0.7$.}
\end{figure}

\subsection{Probability density function for the avalanche size difference}
In recent years, SOC models have been intensively studied
considering time intervals between avalanches in the critical regime \cite{PRL92}.
Here, we follow a different approach which reveals interesting
information on the eventual criticality of the model under
examination. Inspired by recent studies on turbulence and the
time-series of real earthquakes, we introduce the distribution of
returns, i.e., the differences between fluctuation lengths obtained
at consecutive time steps, as $\triangle S(t)=S(t+\delta)-S(t)$, on
the differences between avalanche sizes calculated at time
$t+\delta$ and at time $t$, $\delta$ being a discrete time
interval \cite{PHYSICA309,PRE72,PRE75}.
It should also be noted that, in order to have zero mean, the
returns are normalized by introducing the variable $x$ as:
\begin{eqnarray}
x=\triangle S - <S>
\end{eqnarray}
where $<.>$ stays for the mean value of the given data
set \cite{PRE79}. The signal of the distribution of returns reveals
very interesting results on the criticality of the weighted OFC
model.
In recent works \cite{PRE75,PHYSICA1003,Celikoglu}, it is
shown that the return distributions can be well approximated
by a $q$-Gaussian of type

\begin{eqnarray}
P(x)=P(0)[1-B(1-q)x^{2}]^{1/(1-q)} ,
\end{eqnarray}
(which are the standard distributions obtained in nonextensive statistical
mechanics \cite{Tsallis88,Tsallis2009a} and from where the standard Gaussian
form is obtained as a special case for $q\rightarrow 1$) when the avalanche
size distribution is a power-law with an exponent $\tau$.
Moreover, it is also found that the appropriate $q$ value
could be determined a priori from the exact relation

\begin{equation}
q=\frac{\tau+2}{\tau}
\label{qtau}
\end{equation}
given in \cite{Celikoglu}. It is clear from those efforts
that, as the system size increases, the power-law regime in avalanche size
distribution persists more and more (before arriving the exponential
decay part) which makes the appropriate $q$-Gaussian for the return distribution
to dominate more and more the tails together with the central part.
On the other hand, usually it is very difficult (if not impossible) to reach
very large system sizes (in order to approach thermodynamic limit) in such model
systems. For the small system sizes, considerably short power-law regime is
immediately followed by the exponential decay in avalanche size distribution
and consequently this yields in the return distribution the appropriate
$q$-Gaussian to deteriorate in the tails \cite{Celikoglu,PHYSICA1003}.

In order to explain this tendency, a mathematical simple model for finite-size
effects exhibiting the gradual approach to $q$-Gaussians, has been proposed
using the following differential equation  \cite{Tsallis2009a,2010 J. Phys}:

\begin{equation}
\frac{dy}{d(x^2)}=-b_r y^r - (b_q-b_r) y^q\;\;\;\;(b_q \ge b_r \ge0;\,q>1;\,y(0)=1)\,.
\label{diffeq2}
\end{equation}
For the particular case $r=1$, if one takes $b_1=0$, the solution is given by the
$q$-Gaussian $y=\left[1-(1-q)b_q x^2\right]^{(1/(1-q))} \equiv e_q^{-b_q\,x^2}$.
If $b_q=b_1$, the solution is given by the Gaussian $y=e^{-b_1\,x^2}$.
For the case $b_q>b_1>0$ and $q>1$, we obtain a crossover between these two solutions,
the $|x|\to\infty$ asymptotic one being the Gaussian behavior.
For this particular case with $q>1$, the solution can be found as an
explicit expression of the form $y(x)$, namely,
\begin{equation}
y=\frac{1}{\Bigl[1-\frac{b_q}{b_1}+\frac{b_q}{b_1} \,
e^{(q-1)b_1\,x^2}\Bigr]^{\frac{1}{q-1}}}\, .
\label{diffsol1}
\end{equation}

On the other hand, for the particular case $r=0$ with $q>1$,
the solution can only be given by the explicit $x(y)$ form, namely,
\begin{equation}
x^2=\frac{1}{b_0}\Bigl\{ \\ _2F_1\Bigl[\frac{1}{q},1,1+\frac{1}{q},-\frac{(b_q-b_0)}{b_0} \Bigr] - \,
_2F_1\Bigl[\frac{1}{q},1,1+\frac{1}{q},-\frac{(b_q-b_0)}{b_0} \,y^q \Bigr] y   \Bigr\} \,,
\label{diffsol2}
\end{equation}
where $_2F_1$ is the hypergeometric function.

\begin{figure}[ht] \centering
\scalebox{0.5}{\includegraphics{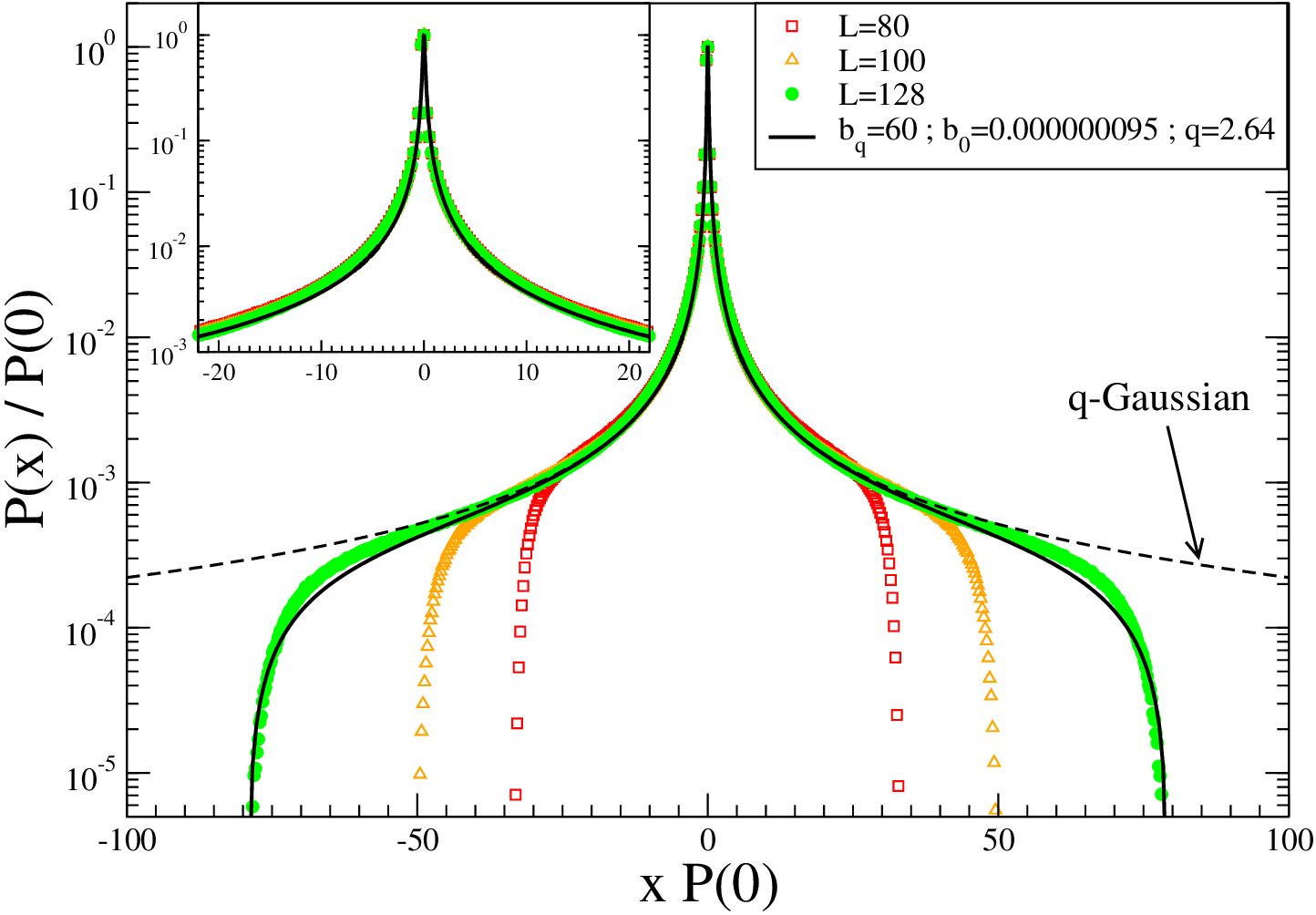}}
\caption{The probability distribution functions of the weighted OFC model
with restricted avalanche definition for representative system sizes.
The crossover formula given in Eq.(\ref{diffsol2}) seems to
describe the tendency in the entire region except the turning points in
the tails (central part is given in the Inset).
As system size increases, it is clearly seen that the return distributions
appear to approach the prefect $q$-Gaussian curve better and better.}
\label{return1}
\end{figure}

Indeed, the weighted OFC model studied here constitutes a very good example to
check the validity of both solutions since (i)~it is an example of a model which
can only be simulated with small system sizes and
(ii)~the model allows us to define two types of avalanche definition, one of
which seems to produce return distributions that can be approached by Eq.(\ref{diffsol1}),
whereas the other definition yields return distributions that can be given
by Eq.(\ref{diffsol2}). The standard way of defining an avalanche (the one also
used throughout this work) is to include each triggered site only once during an
avalanche which restricts the size of an avalanche with the size of the system.
This definition results in the return distributions shown in Fig.~\ref{return1}.
It is easily seen that the returns are well approximated by the crossover formula given in
Eq.(\ref{diffsol2}) with $q=2.64$ which comes a priori from Eq.(\ref{qtau}).
It is also evident from this figure that, as $L\rightarrow\infty$ limit
is approached, it seems that the return distributions would converge to the
$q$-Gaussian with $q=2.64$ for the entire region including the central part
and the tails.

\begin{figure}[ht] \centering
\scalebox{0.5}{\includegraphics{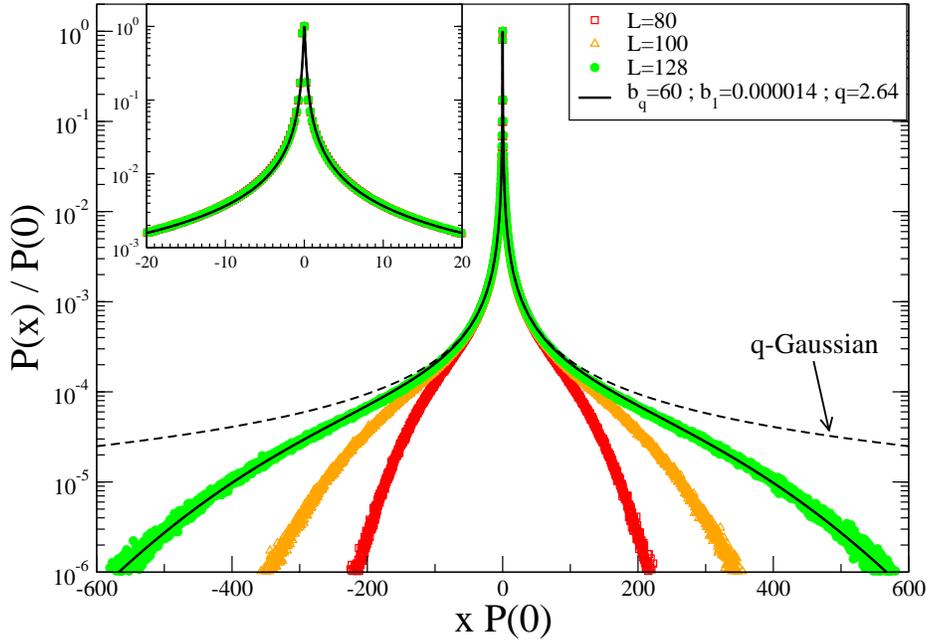}}
\caption{The probability distribution functions of the weighted OFC model
with non-restricted avalanche definition for representative system sizes.
The crossover formula given in Eq.(\ref{diffsol1}) seems to perfectly
describe the tendency in the entire region (central part is given in the Inset).
As system size increases, it is clearly seen that the return distributions
appear to approach the perfect $q$-Gaussian curve better and better.}
\label{return2}
\end{figure}

Another way of defining an avalanche is to relax the restriction that allows
each site to trigger only once during an avalanche. This means that, during a
running avalanche, one site can be triggered more than once which clearly
relaxes the restriction of having maximum avalanche sizes of the order of
system size. The use of such definition does not change the value of the
avalanche size distribution exponent $\tau$ but results in a smoother crossover
from the power-law regime to exponential decay part. This observed tendency
would be expected to have an effect also in the return distributions.
This can be seen in Fig.~\ref{return2} where the return distributions can
now be well approached by the crossover formula given in Eq.(\ref{diffsol1}).
The gradual approach to the perfect $q$-Gaussian is evident as the system
size is increased. It is also worth noting here that the observed behavior of
return distributions do not depend on the interval $\delta$ considered for the
avalanche size difference, which is also the case for the worldwide and the
northern California catalogs \cite{PRE75}.

As a result, one can conclude here that the behavior of the return
distributions is very different from a Gaussian shape and seems to
be well approached by one of the two crossover formulas
(either by Eq.(\ref{diffsol1}) if the non-restricted definition of
avalanche is used or by Eq.(\ref{diffsol2}) if the restricted definition
of avalanche is used). As far as we know, this constitutes the first
example in literature where the two forms of these crossover solutions
can be used together in the same model system.
Finally, all the numerical findings obtained here suggest that,
as the thermodynamic limit is approached, the behavior of the return
distributions seems to converge to the appropriate $q$-Gaussian shape
in the entire region.

\subsection{Statistics of waiting time in weighted OFC model}
Recently, a new necessary SOC signature has been proposed, in the
context of solar flare dynamics. It is based on a different type of
statistics that deals with waiting times, i.e., the time intervals
between two successive bursts or avalanches. In this section, we
consider the waiting time of avalanche sequences generated by our
weighted model. The waiting time is defined as the time between the
first trigger and the second one. We use the overall statistical
method and statistics of the waiting times for any node, then the
distribution of all nodes. After this, we calculated probability
distribution of waiting times. It can be argued that, if the triggers
are not correlated, the process should be somehow related to a
Poisson process, and the probability distribution function of the
waiting times should be an exponential law.
However,the existence of extended power laws in the waiting-time
probability distribution function of solar flare measurements
has been noticed by several authors \cite{PRL83,Astrophys509}.
Several years ago, Christensen et al. showed that waiting times
would follow power-law distributions if only events larger than a
certain size are considered in the context of a spring-block
model for earthquakes \cite{Geophys97}. In this paper, we also analyze
the waiting time distribution of the weighted OFC model which exhibits
a clear nonexponential behavior as can be seen in Fig.~6.
The power-law regime of the waiting time distribution lasts about two decades.

\begin{figure}[ht] \centering
\scalebox{1.2}{\includegraphics{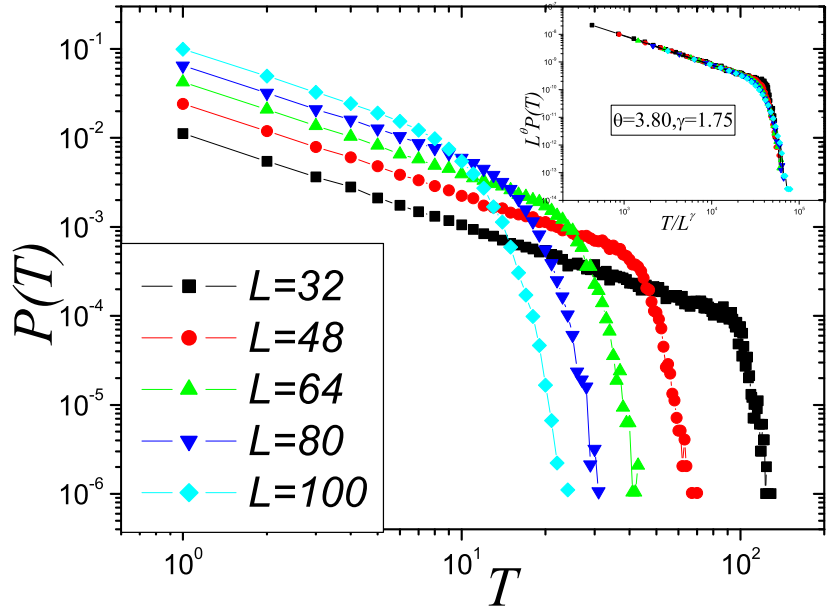}}
\caption{The waiting time distribution and its scaling analysis.}
\end{figure}
We propose a scaling relation for the waiting-time distribution
of the form

\begin{eqnarray}
P(T)&\propto& L^{-\theta}g(T/L^{\gamma}) \,\, ,
\end{eqnarray}
with the scaling exponents $\theta=3.80$ and $\gamma=1.75$, which
are shown (see the inset of Fig.~6) to be consistent with the data
coming from our model.

\section{Summary and Conclusion}
In order to obtain the inhomogeneous network and different local
friction and elasticity, we have introduced the weighted edge to
improve the original redistribution rule. We have shown
self-organized criticality in the weighted coupled map lattice.
The probability density functions of the avalanche size differences
(namely, return distributions) appear to exhibit fat tails that can be
approached by a $q$-Gaussian shape, in the thermodynamic limit,
with an appropriate value of $q$ coming a priori from the avalanche size
exponent $\tau$.
Moreover, for the small system sizes, the observed behavior of the
return distributions seems to obey the crossover formula proposed
in \cite{2010 J. Phys} in order to explain the transition from the
$q$-Gaussian behavior to the Gaussian observed so far in some other
model systems with small system sizes.
These results could be interpreted that there are no correlations
between any two seismic behavior. Our findings support
the hypothesis that even the statistical data of previous
earthquake is known, the magnitude of the next earthquake is
still unpredictable.
Finally, the scaling relation of waiting times for the weighted
OFC model has been discussed and obtained.

\section{ACKNOWLEDGEMENTS}
This work has been supported by the National Natural Science
Foundation of China under Grant No.10675060 and
by Ege University under the Research Project number 2009FEN027.
We thanks C.P.Zhu and H.Kong for useful discussions.

\begin{center}

\end{center}
\end{document}